# Correlation invariance unlocks robust calibration-free orbital-angular-momentum multiplexing transmission under dynamic scattering scenarios

**Haoran Li,[a,b,†] Zhiyuan Wang,[a,b,†] Zhipeng Yu,[a,b,d,†] Xingpeng Du,[c] Tianting Zhong,[a,b] Jixiong Pu,[c] Ziyang Chen,[c] Vinu R V,[c,*] Xiangping Li,[d,*] and Puxiang Lai,[a,b,e,*]**

[a] Department of Biomedical Engineering, The Hong Kong Polytechnic University, Hong Kong, China
[b] Shenzhen Research Institute, The Hong Kong Polytechnic University, Shenzhen, China
[c] College of Information Science and Engineering, Huaqiao University, Xiamen, China
[d] Guangdong Provincial Key Laboratory of Optical Fiber Sensing and Communications, Institute of Photonics Technology, Jinan University, Guangzhou, China
[e] Photonics Research Institute, Hong Kong Polytechnic University, Hong Kong, China

[†] These authors contributed equally to the work.
[*] Correspondence to Vinu R V at vinurv@hqu.edu.cn, Xiangping Li at xiangpingli@jnu.edu.cn, and Puxiang Lai at puxiang.lai@polyu.edu.hk

**Abstract**. Orbital angular momentum (OAM) multiplexing offers a promising approach to high-capacity optical communication by harnessing the orthogonality of vortex beams. However, its practical deployment is severely limited in real-world settings where dynamic scattering media, such as turbulent atmosphere, distort multiplexed fields into random speckles and disrupt OAM demultiplexing. Although existing wavefront shaping and deep learning methods can mitigate static distortions, they fail under time-varying scattering conditions, leading to significant crosstalk and unreliable recovery. Here, we introduce a new concept, correlation invariance, which enables scattering-immune, robust OAM multiplexed transmission through dynamic media. By capturing orthogonally polarized speckle holograms in a compact common-path geometry and computing their intensity cross-correlation, dynamically imposed scattering phases are cancelled out while deterministic object information is preserved. This allows single-shot reconstruction of both amplitude and phase of the input OAM-multiplexed fields, without any pre-calibration or training. As a proof of principle, we demonstrate high-fidelity transmission of 24-bit RGB data with 99.61% accuracy under static scattering and 98.97% accuracy under dynamic scattering. This approach addresses a long-standing barrier in OAM-based systems and opens avenues for robust high-capacity optical communications, encryption, and imaging in dynamic scattering environments.

**Keywords**: Correlation invariance, Orbital angular momentum, OAM multiplexing, Scattering, Speckle correlation holography

# 1 Introduction

Vortex beams have garnered significant attention in recent years due to their helical wavefronts and unique physical characteristics [1]. Characterized by the phase component $exp(il\varphi)$, where $\varphi$ represents the azimuthal angle and $l$ denotes the topological charge (TC), these beams carry orbital angular momentum (OAM) [2-4], rendering them highly valuable for multidisciplinary potentials [5-14]. In optical communications, OAM multiplexing exploits the orthogonality of distinct OAM modes to enable high-capacity data transmission, offering a promising route towards channel capacity expansion [9-12]. A critical requirement for such systems is the precise decomposition of the OAM spectrum at the receiver end, which can be accomplished with high fidelity in free space. However, OAM-multiplexed beams are inevitably affected by scattering when transmitted over long distances or in complex environments [15-17]. Conventional OAM spectrum decomposition methods [11, 18-20] encounter significant challenges in such scattering environments, as OAM-multiplexed beams are distorted into random speckle patterns after propagating through a scattering medium, thereby preventing the recovery of the original field [17, 21-25]. This fundamental limitation severely restricts the practical deployment of OAM-multiplexing-based optical communication in real-world environments.

Several strategies have been proposed to recover OAM information through scattering media. Transmission matrix (TM) methods [26-28] are utilized to reconstruct the multiplexed fields from speckle patterns [29-33], but they require system pre-calibration and assume medium stationarity, posing critical challenges for dynamic scattering environments. Deep learning approaches circumvents TM measurement yet demand large training datasets and degrade when the scattering environment changes [34-38]. Speckle correlations strategies exploit correlations between speckle patterns of known and unknown OAM modes [39, 40] but need multiple measurements (≥2 per OAM mode) and a static medium state during acquisition. While recent work has demonstrated single-OAM detection in dynamic scattering media using multiple detectors [41], high-capacity OAM-multiplexed transmission under realistic, time-varying scattering conditions remains unrealized. Critically, such a capability would significantly broaden the realm of OAM multiplexing.

Here, we introduce a new concept, correlation invariance, and implement it via a common-path dual-polarization speckle correlation holography (CPDP-SCH) technique to achieve single-shot reconstruction of OAM-multiplexed fields through dynamic scattering media. By capturing orthogonally polarized speckle holograms that share an identical optical path and computing their intensity cross-correlation, the random, time-varying scattering phase is inherently canceled, while deterministic object information is preserved. This allows recovery of both amplitude and phase directly from a single measurement without pre-calibration, phase shifting, or iterative algorithms. We demonstrate robust, near-perfect transmission of 24-bit RGB data with 99.61% accuracy under static scattering and 98.97% accuracy under dynamic scattering, overcoming a long-standing barrier to OAM-based applications in real-world scattering environments.

## 2 Methods

### 2.1 Principle of correlation invariance

Fig. 1a illustrates a simplified schematic of CPDP-SCH. The full area (1920×1080 pixels) of a spatial light modulator (SLM) is loaded with a customized phase pattern that allows independent generation of beams in different sections, each assignable to distinct polarization directions (inset of Fig. 1a). In this study, four regions are defined to generate the target OAM field, a plane wave, and two lens phase fields. Through a beam splitter and polarization optics, the plane wave and one of the lens fields are converted to vertical polarization and combined with the OAM field and the other lens field, forming a composite beam (details in Section 2.3). Within this beam, the horizontally polarized OAM component and the vertically polarized plane wave maintain precise spatial overlap during propagation. After passing through the dynamic scattering medium, they produce polarization-specific speckle patterns on the detection plane. Simultaneously, the two lens phase fields converge identically onto the same spot on the scattering medium surface and, after diffraction, serve as off-axis speckle reference beams for their representative polarization channels. Consequently, two orthogonally polarized off-axis speckle holograms are generated and captured simultaneously in a single frame by a polarization camera. Crucially, throughout the entire transmission process, all fields, including the OAM field, the plane wave, and the two lens phase fields, share a common optical path.

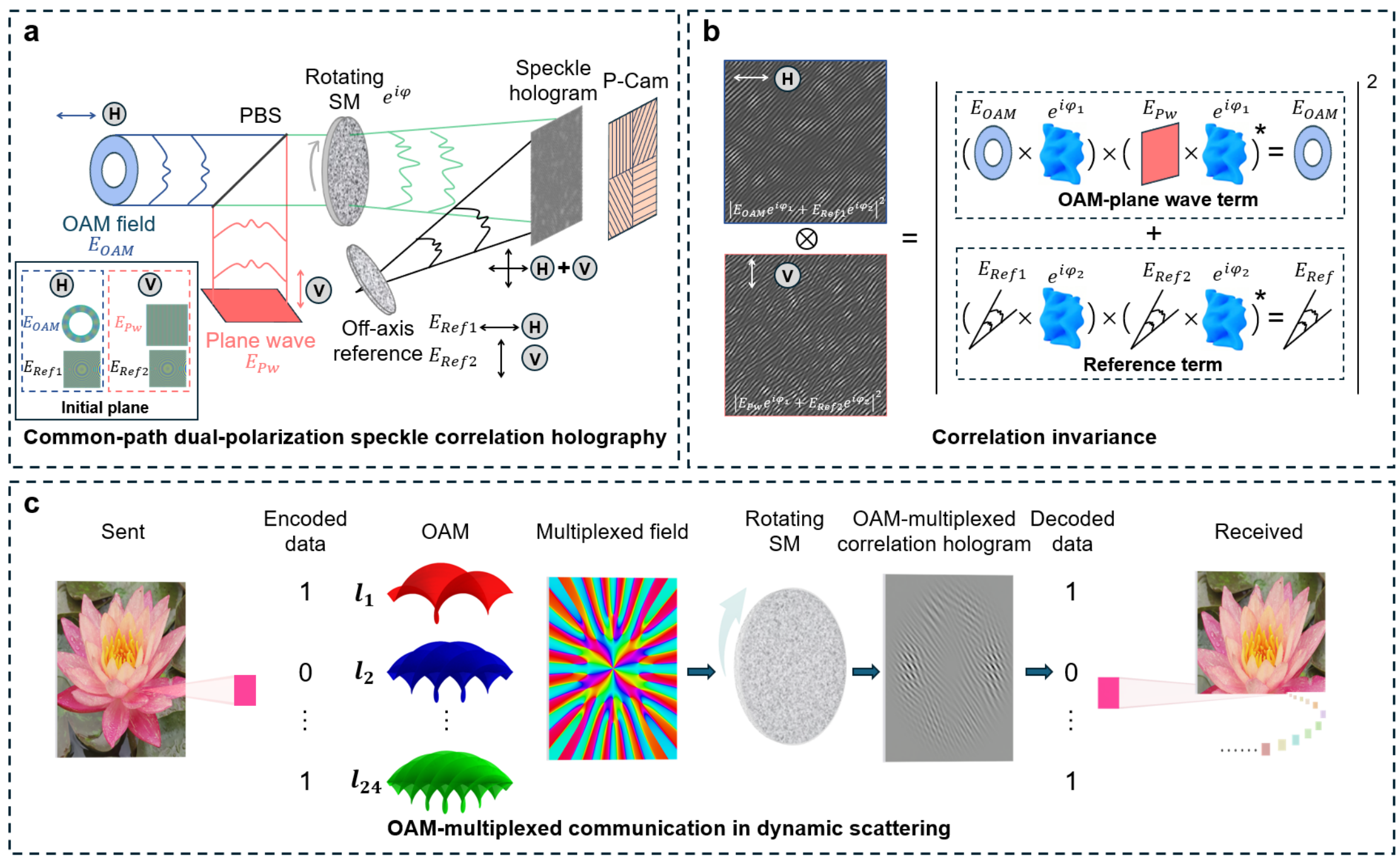

**Fig. 1 Concept and principle of common-path dual-polarization speckle correlation holography (CPDP-SCH). a.** Schematic of the CPDP-SCH system. The SLM generates an OAM field, a plane wave, and two lens-phase reference fields via a customized phase pattern (inset). The plane wave is converted to vertical polarization and combined with the horizontally polarized OAM field. All beams share a common path through the dynamic scattering medium (SM). Orthogonally polarized off-axis speckle holograms are simultaneously recorded by a polarization camera (P-Cam). **b.** Principle of correlation invariance in scattering media. Cross-correlation of polarization-separated speckle holograms cancels the random scattering phases while preserving deterministic OAM information, enabling full-field recovery. **c.** OAM-multiplexed communication through dynamic scattering. A color image is encoded pixel-wise into 24 OAM channels, transmitted through the scattering medium, and decoded via CPDP-SCH to reconstruct the received image. Abbreviations: PBS, polarizing beam splitter; SM, scattering medium; Ref, reference beam; Pw, plane wave; P-Cam, polarization camera.

Fig. 1b illustrates the principle of correlation invariance. When an OAM field and a plane wave co-propagate through the same scattering medium, they are subjected to an identical random phase modulation and each generates its own polarization-specific speckle pattern on the detection plane. By interfering each with the same off-axis reference speckle field, two speckle holograms are formed, recorded, and separated by the polarization camera, as shown in the left panel of Fig. 1b. The cross-correlation of these two holograms yields an expression proportional to the squared modulus of two terms: an OAM-plane wave term and a reference term (details in Section 2.2). In the OAM-plane wave term, the random scattering phase cancels via multiplication with its complex conjugate, leaving the product of the OAM field and the plane wave, which preserves the deterministic OAM information, thereby ensuring the invariance of the input field. Similarly, the random phase in the reference term is also removed by cross-correlation, extracting the off-axis reference field in this expression. This recovered off-axis reference field then interferes with the OAM term, producing a correlation hologram that retains full amplitude and phase information of the input light. Thus, correlation invariance effectively cancels the scattering-induced phases distortions while preserving the deterministic object information.

Fig. 1c outlines the OAM-multiplexed communication process in a dynamic scattering environment using the CPDP-SCH method. A color image is transmitted pixel by pixel, with each pixel's data decomposed into three RGB channels and combined into a 24-bit binary code. This code is encoded using 24 distinct OAM beams, each assigned a specific topological charge. The corresponding encoded phase pattern is loaded onto an appropriate region of the SLM, and the resulting field propagates through the dynamic scattering medium. The polarization camera records the orthogonally polarized speckle holograms in a single shot. Cross-correlation of these holograms yields the correlation hologram corresponding to the transmitted OAM-multiplexed field, from which the original OAM-multiplexed field is reconstructed. Finally, OAM power spectrum analysis decodes the data, recovering the color pixel information. This process is repeated sequentially until all pixels of the target color image are transmitted.

### 2.2 Derivation of the correlation invariance

In the implemented system, the composite field after the SLM in the output modulation plane contains four components: an OAM-encoded field, a plane wave, and two lens phase fields, each associated with a specific polarization. After propagating through the scattering medium and a distance of $Z_0$, polarization-separated off-axis speckle holograms are formed on the detection plane. The field propagation from the scattering medium to the detection plane can be expressed as:

$$\begin{cases} E_{(R)H}(r_1) = \int h_{Z_0}(r_1, \rho_1)\left[E_{(R)H}(\rho_1)e^{i\varphi(\rho_1)}\right] d\rho_1 \\ E_{(R)V}(r_2) = \int h_{Z_0}(r_2, \rho_2)\left[E_{(R)H}(\rho_2)e^{i\varphi(\rho_2)}\right] d\rho_2 \end{cases}, \tag{1}$$

where $r$ and $\rho$ denote the coordinates on the detection and the scattering planes, respectively; subscripts 1 and 2 attached to $r$ and $\rho$ are used to distinguish the horizontal and vertical polarization channels; $h_{Z_0}(r_1, \rho_1)$ is the free-space diffraction kernel between the scattering and the detection planes; $\varphi(\rho)$ is the random phase imposed by the scattering medium. Here, $E_{(R)H}(r_1)$ and $E_{(R)V}(r_2)$ represent the speckle fields in the two polarization channels, encompassing two specific cases: the object speckle patterns $E_H(r_1)$ and $E_V(r_2)$ generated by the OAM-encoded beam and plane wave, and the reference speckle patterns $E_{RH}(r_1)$ and $E_{RV}(r_2)$ formed by the lens-modulated fields; the latter serves as the off-axis reference speckles for forming the corresponding holograms. Consequently, the polarization-specific off-axis speckle holograms $U_H(r_1) = E_H(r_1) + E_{RH}(r_1)$ and $U_V(r_2) = E_V(r_2) + E_{RV}(r_2)$ will be generated in the detection plane.

The polarization camera captures the intensities of these two fields: $I_H(r_1) = |U_H(r_1)|^2$ and $I_V(r_2) = |U_V(r_2)|^2$, which is illustrated in Fig. 1b.

For a random process obeying Gaussian statistics, the cross-correlation of the intensity fluctuations of these two captured holograms, $C(r_1, r_2)$, is given by:

$$C(r_1, r_2) = \langle \Delta I_H(r_1) \Delta I_V(r_2) \rangle = \left| c_{H,V}(r_1, r_2) \right|^2, \tag{2}$$

where $< \cdot >$ denotes the ensemble average and $\Delta$ represents the intensity fluctuation relative to the mean. $c_{H,V}(r_1, r_2)$ is the field cross-correlation function expressed by:

$$\begin{aligned} \left|c_{H,V}(r_1, r_2)\right|^2 &= \left|\left\langle \left(E_H(r_1) + E_{RH}(r_1)\right)\left(E_V(r_2) + E_{RV}(r_2)\right)^* \right\rangle\right|^2 \\ &= |\langle E_H(r_1)E_V^*(r_2) + E_{RH}(r_1)E_V^*(r_2) + E_H(r_1)E_{RV}^*(r_2) + E_{RH}(r_1)E_{RV}^*(r_2)\rangle|^2 \\ &= |\langle E_H(r_1)E_V^*(r_2) + E_{RH}(r_1)E_{RV}^*(r_2)\rangle|^2 \\ &= |\langle E_H(r_1)E_V^*(r_2)\rangle + \langle E_{RH}(r_1)E_{RV}^*(r_2)\rangle|^2. \end{aligned} \quad (3)$$

In Equation (3), the cross-terms $\langle E_{RH}(r_1)E_V^*(r_2)\rangle = 0$ and $\langle E_H(r_1)E_{RV}^*(r_2)\rangle = 0$ because the component in the pairs $E_{RH}(r_1), E_V^*(r_2)$ and $E_H(r_1), E_{RV}^*(r_2)$ originate from independent random modulations occurring in distinct regions of the scattering medium. Notably, the speckle field $E_H(r_1)$, generated by the input vortex beam or OAM-multiplexed beam after transmission through the scattering medium, is encoded within the final expression of $c_{H,V}(r_1, r_2)$.

Substituting $r + \Delta r$ for $r_1$ and $r$ for $r_2$, $C(r_1, r_2)$ transforms to:

$$C(r_1, r_2) = |\langle E_H(r + \Delta r)E_V^*(r)\rangle + \langle E_{RH}(r + \Delta r)E_{RV}^*(r)\rangle|^2. \quad (4)$$

Substituting Equations S(1) and S(3) in Supplementary Note 1 into Equation (4), and assuming $\rho_1 = \rho_2 = \rho$ at the scattering medium plane due to the corresponding fields have precise spatial overlap, the random phases factor $\varphi(\rho_1)$ and $\varphi(\rho_2)$ imposed by the scattering medium canceled in the integral, and the Equation (4) is simplified to (see Supplementary Note 1 for more details):

$$C(r_1, r_2) = \left|c_{H,V}^{(O)}(\Delta r) + c_{H,V}^{(R)}(\Delta r)\right|^2, \quad (5)$$

with

$$\begin{cases} c_{H,V}^{(O)}(\Delta r) = \dfrac{1}{\lambda^2 Z_0^2}\displaystyle\int \left[\sum_l E_{H,l}^{O(f)}(\rho)\right] exp\left(\dfrac{-ik(\Delta r)\rho}{Z_0}\right) d\rho \\ c_{H,V}^{(R)}(\Delta r) = \dfrac{1}{\lambda^2 Z_0^2}\displaystyle\int circ\left(\dfrac{\rho - \rho_s}{a}\right) exp\left(\dfrac{-ik(\Delta r)\rho}{Z_0}\right) d\rho \end{cases}, \quad (6)$$

where $\sum_l E_{H,l}^{O(f)}(\rho)$ denotes the OAM-encoded field at the scattering medium plane, which can be expressed as the superposition of the fields resulting from vortex beams that originate from the modulation output plane and propagate over a distance $f$; $l$ is the TC, $\omega_0$ is the beam waist, and $\varphi$ is the azimuthal angle; $circ\left(\frac{\rho-\rho_s}{a}\right)$ is the circular apertures of diameter $a$ generated by lens-phase fields focus to identical positions $\rho_s$ on the scattering medium plane; $\lambda$ is the wavelength of the incident light, and $k = \frac{2\pi}{\lambda}$ is the wavevector.

Thus, the cross-correlation of the intensity fluctuations, computed from the off-axis speckle holograms ($I_H$ and $I_V$) of the two polarization states, yields a correlation hologram encoding the complete information of the OAM-encoded beam. Unlike conventional speckle correlation imaging, CPDP-SCH employs the complex correlation function for holographic recording and reconstruction, enabling simultaneous restoration of amplitude and phase components. In Equation (6), the complex correlation $c_{H,V}^{(O)}(\Delta r)$ encodes the information of single vortex beam or OAM-multiplexed beam, while the reference complex correlation $c_{H,V}^{(R)}(\Delta r)$ generates the off-axis fringes. Also, the off-axis parameters can be flexibly controlled through the design of the loaded phase patterns (see Supplementary Note 4 for more details). Furthermore, the original OAM-encoded information can be efficiently retrieved by applying a Fourier transform-based digital analysis combined with spectral filtering [42-44] (see Supplementary Note 3 for detailed processing flowchart). This derivation confirms that the proposed CPDP-SCH enables single-shot recovery of OAM-multiplexed fields through scattering media.

### 2.3 Experimental Setup

Fig. 2a shows the experimental setup for implementing CPDP-SCH. A 632 nm wavelength beam emitted from a He-Ne laser (25-LHP-928-230, Melles Griot) is expanded via a beam expander to fully illuminate the surface of a SLM (PLUTO-VIS, Holoeye). After passing through a horizontally oriented polarizer (P) and a beam splitter (BS1), the beam is modulated by the SLM according to the loaded phase pattern, which encodes a grating pattern to realize complex amplitude modulation [45]. As depicted in Fig. 2b, this phase pattern is encoded by combining an OAM field, a plane wave, and a lens phase field to realize the desired complex amplitude modulation. Fig. 2c illustrates how the phase pattern generates the desired complex field. The modulated beam is reflected by BS1 and focused by lens L1, generating multiple diffraction orders at its focal plane. A pinhole isolates the +1 diffraction order, which is subsequently split by BS2 into two paths: one remains horizontally polarized, while the other is converted to vertical polarization by a half-wave plate (HWP). These two orthogonally polarized beams are reflected by two mirrors (M2 and M3) and recombined through a polarized beam splitter (PBS). Careful alignments of mirrors M2, M3, and the PBS ensure spatial overlap between the OAM-encoded beam (horizontal polarization) and the plane wave (vertical polarization), while the lens-phase-modulated beams in both paths also coincide spatially (see Supplementary Note 2 for more details). Lenses L1 and L2 form a 4f system, producing a complex amplitude-modulated beam at the back focal plane of L2, which is filtered by an aperture to retain the final combined output. From the perspective of polarization state, the combined beam is divided into two components: the horizontally polarized light contains an OAM-encoded beam (upper left) and lens-phase-modulated beam (lower right), whereas the vertically polarized light comprises a plane wave (upper left) and another lens-phase-modulated beam (lower right). After propagating over the designated focal length, the lens-phase-modulated beams focus onto the surface of a ground glass (GG, 220 grits), generating off-axis reference speckle fields in both polarizations via diffraction. Simultaneously, the OAM-encoded beam and plane wave produce corresponding polarization-specific speckle patterns. This results in polarization-separated off-axis speckle holograms—OAM-encoded beam holograms in horizontal polarization and plane wave holograms in vertical polarization—which are captured by the polarization camera. Since each pixel of the polarization camera integrates a micro-polarizer array, the speckle patterns in four different polarization directions (0°, 45°, 90°, and 135°) could be simultaneously obtained. Final reconstruction of the OAM-encoded field is achieved through cross-correlation and computational processing.

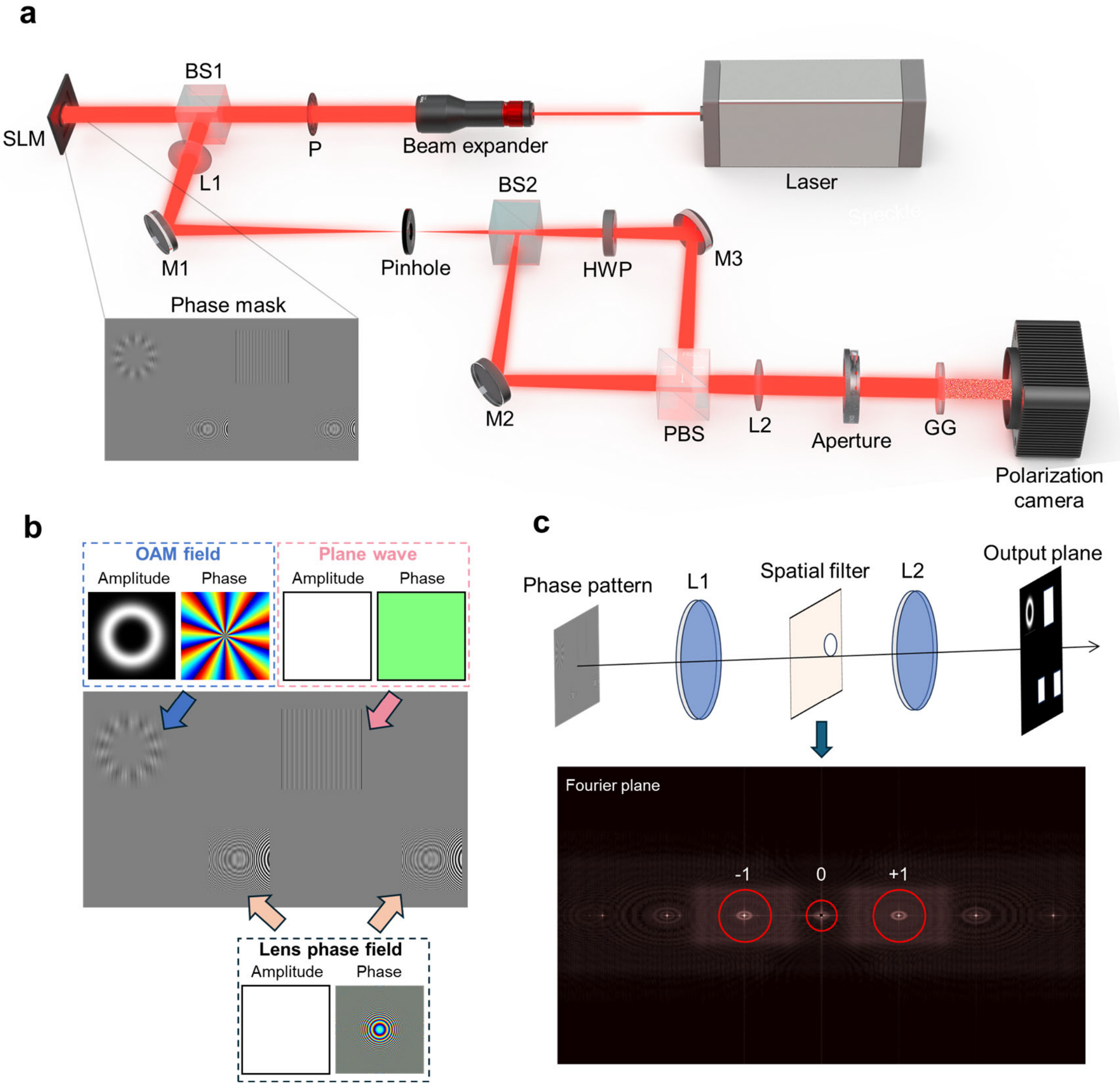

**Fig. 2**. **Experimental implementation of CPDP-SCH and complex amplitude modulation strategy.** **a**, Optical layout. A 632 nm He-Ne laser beam is expanded, polarized (P), and split (BS1). The SLM modulates light via a specific grating-encoded pattern. The +1-diffraction order (isolated by L1 and pinhole) is split (by BS2) into Path 1 (H-pol) including OAM-encoded field plus lens phase field and Path 2 (V-pol) including plane wave plus lens phase field (converted via HWP). Mirrors (M2, M3) and polarizing beam splitter (PBS) recombine beams with spatial overlap. A 4f system (L1, L2) delivers the combined field to the scattering medium (GG). Lens-modulated fields focus onto GG, generating polarization-separated off-axis reference speckles. Polarization camera simultaneously captures holograms at 0°, 45°, 90°, and 135°. Abbreviations: P, polarizer; BS, beam splitter; SLM, spatial light modulator; L, lens; M, mirror; HWP, half-wave plate; PBS, polarizing beam splitter; GG, ground glass. **b**, Phase pattern design. Different regions of the SLM pattern are encoded by different beams to achieve the generation of corresponding beams. **c**, Generation of the complex optical field. The modulated field

is Fourier-transformed by L1; a spatial filter selects the +1 diffraction order to generate the desired complex field in the output plane.

## 3 Results

### 3.1 Single OAM reconstruction

We first validated the performance of CPDP-SCH by reconstructing single vortex beams transmitted through a static scattering medium. Vortex beams with TCs of 1, 8, 16, and 24 served as test inputs. Figs. 3a and 3b show the polarization-separated speckle patterns ($I_V$, $I_H$) captured for TC=1. The central regions of the corresponding correlation holograms, obtained by cross-correlation of orthogonal speckle patterns, exhibit clear off-axis structures (Figs. 3c-f). These fringes correspond to the interference in free space between the incident vortex beam and the off-axis reference beam, confirming that each correlation hologram encodes the complete complex field the input vortex beam. Subsequent Fourier analysis reconstructed amplitude (Figs. 3g-j) and phase (Figs. 3k-n) distributions for each vortex beam. Phase dislocations appearing within azimuthal periods clearly identified each TC, and the reconstructed profiles show excellent agreement with the input OAM characteristics. These results demonstrate that CPDP-SCH achieves high-fidelity OAM information restoration through scattering media.

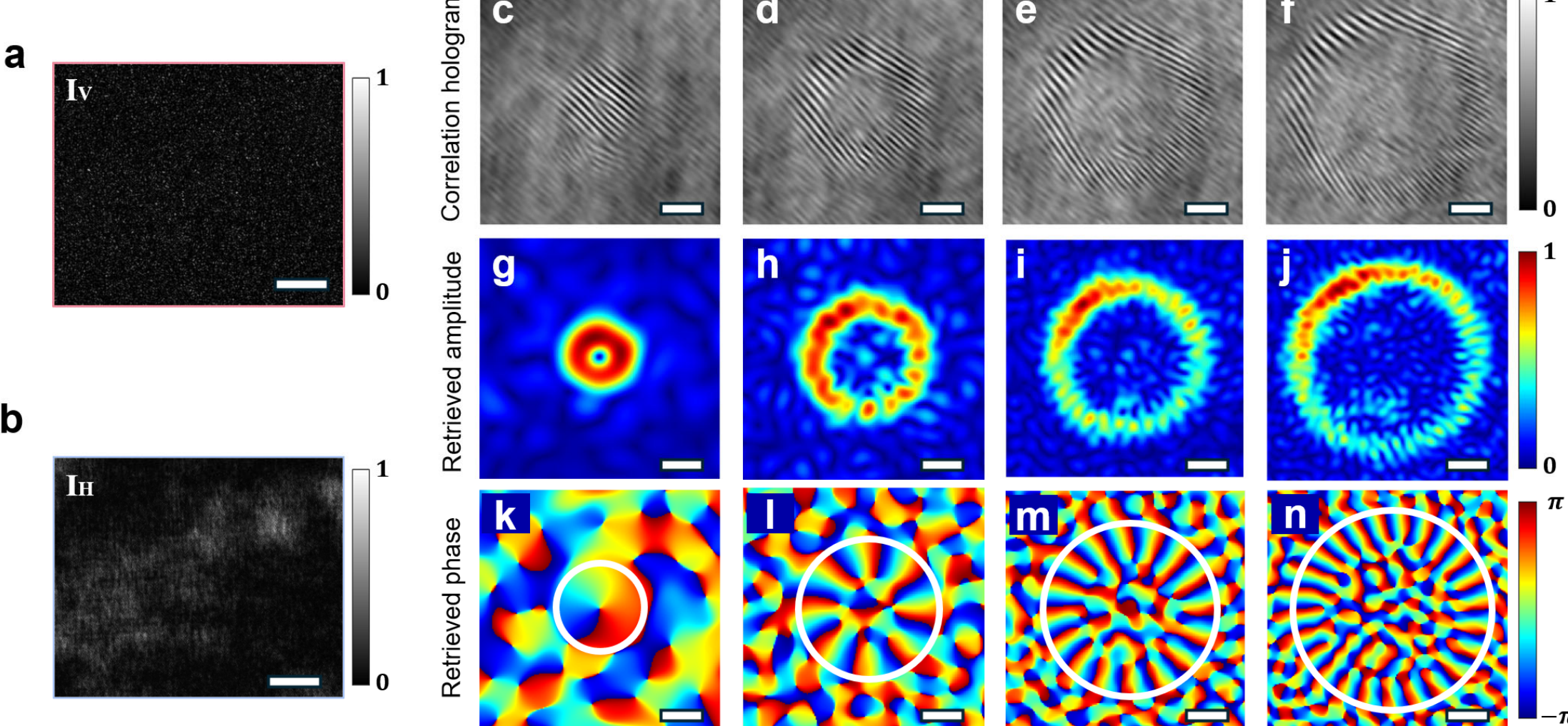

**Fig. 3. Single vortex beam reconstruction via CPDP-SCH in static scattering media. a-b**, Experimentally captured off-axis speckle holograms for horizontal ($I_H$) and vertical ($I_V$) polarizations, generated by vortex beam with topological charge TC=1. **c-f** Central regions of the corresponding

correlation holograms for TC=1, 8, 16, and 24, respectively, exhibiting distinct off-axis fringe structures. **g–j**, Reconstructed amplitude profiles. **k–n**, Reconstructed phase distributions with characteristic helical dislocations confirming each TC value. Scale bars: 800 μm (a-b), 400 μm (g-n).

### 3.2 OAM-multiplexed information reconstruction

Next, we validated CPDP-SCH for OAM-multiplexed data transmission through static scattering media. For a reconstructed OAM-multiplexed field $E_r$, spatial mode decomposition is required to obtain its constituent OAM components. This was accomplished by calculating its power spectrum on the corresponding OAM basis. The process began with computing the complex-valued coefficients $c_n$ through the integral of $E_r$ with each vortex beam component in the OAM basis in polar coordinates:

$$c_n = \iint E_r E_{ln}^* \, r dr d\phi, \tag{7}$$

where $E_{ln}$ is the vortex beam corresponding to the $n$th OAM mode, $n$ denotes the index of the beam within the OAM basis, $r$ and $\phi$ represent the radial distance and azimuthal angle in polar coordinates, respectively, and the superscript $*$ denotes the complex conjugate operation. Subsequently, the power spectrum $P(l_n)$ of $E_r$ on the OAM basis is obtained:

$$P(l_n) = \frac{|c_n|^2}{\sum_{n=1}^{N_l} |c_n|^2}, \tag{8}$$

where $N_l$ is the number of OAM modes in the basis. Analyzing $P(l_n)$ reveals the composition of the individual OAM component within the multiplexed light field $E_r$, thereby enabling information transmission applications.

We first examined the orthogonality between the retrieved vortex beams to determine the optimal selection of OAM basis for achieving OAM-multiplexing. By calculating the power spectrum corresponding to each reconstructed vortex beam, we assessed the orthogonality within the basis and quantified the level of crosstalk. Fig. 4a presents the power spectra for the experimental OAM bases with TC ranging from -12 to 12 (step size 1) and from -24 to 24 (step size 2), respectively. The maximum crosstalk for single vortex beams transmission was -11.42 dB for step size of 1 and -13.38 dB for the step size 2, confirming excellent orthogonality in both cases. Statistical analysis (right panel, Fig. 4a) yielded average power spectrum values of 0.955 and 0.971, corresponding to average crosstalk levels of -13.46 dB and -15.37 dB, for step sizes 1 and 2, respectively. The error bars indicate the maximum and minimum transmission power spectrum values within the respective OAM basis.

Given its lower crosstalk, the step-size-2 OAM basis was employed to encode 24-bit RGB information for color image transmission (Fig. 4b), where 24 vortex beams were combined via OAM-multiplexing to represent 24-bit RGB information. Each color channel (R, G, B) was decomposed into three independent 8-bit binary channels, encoded by three groups of vortex beams with TCs ranges [-24, -10], [-8, 8], and [10, 24], respectively. Each vortex beam was multiplied by its corresponding binary digit (0 or 1), and the results were summed to generate the final multiplexed input. The right panel of Figure 4b shows an example multiplexed input for encoding a color pixel alongside its corresponding CPDP-SCH reconstruction. OAM mode decomposition of the recovered field yields a power spectrum that matches the encoded OAM components (Fig. 4c). Using the thresholding method described in Ref. [29], the OAM mode component present in the retrieved multiplexed field could be identified, enabling independent processing of the R, G, and B channels for accurate recovery of the encoded color information.

Beyond single-pixel transmission, we also successfully transmitted a 100×100-pixel color image (Fig. 4d) via OAM multiplexing combined with CPDP-SCH. The reconstructed image was recovered with a near-perfect error rate of 0.039 %, demonstrating high-fidelity OAM-multiplexed data transmission through scattering media.

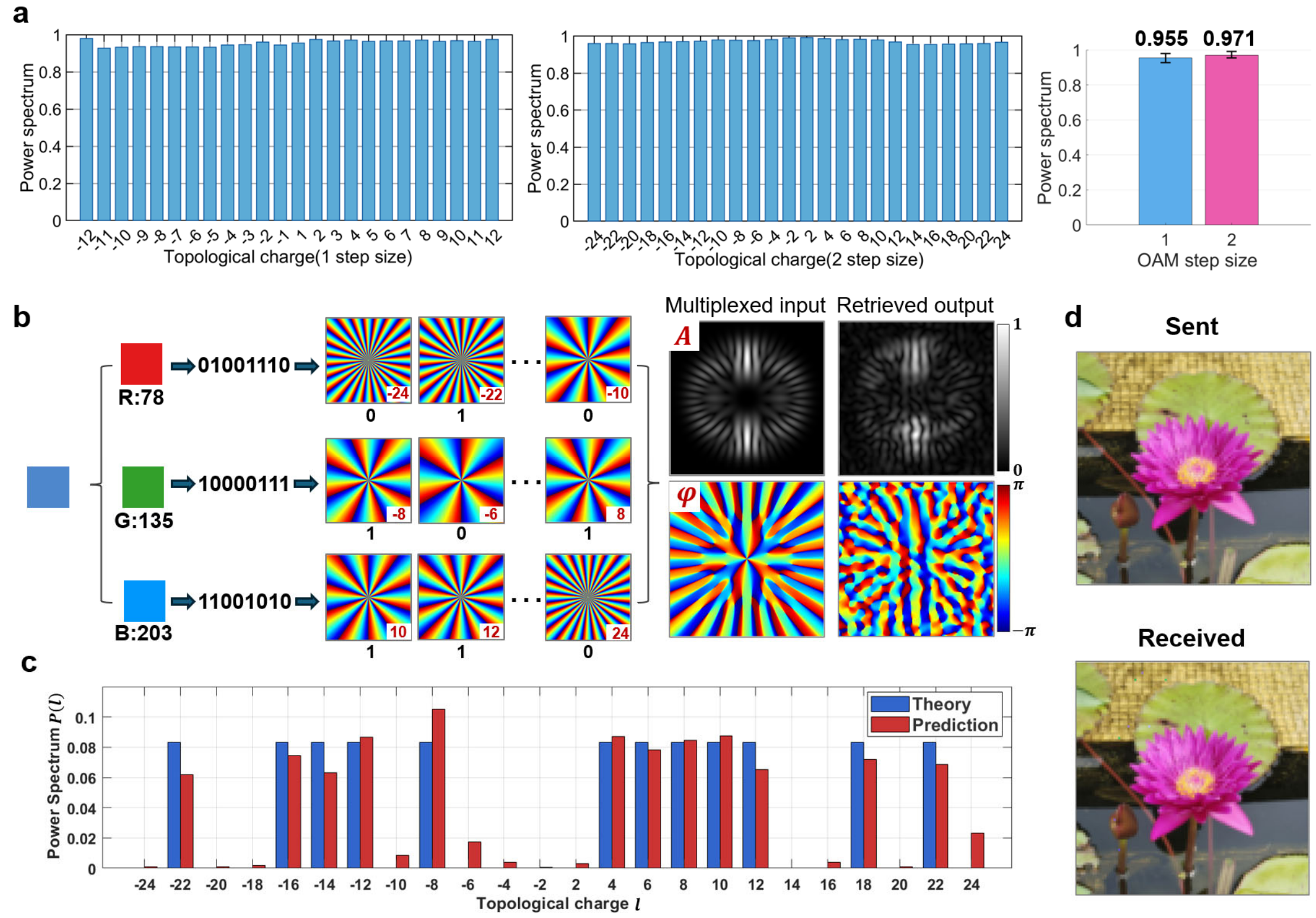

**Fig. 4**. **OAM-multiplexed data transmission through static scattering media. a**, Left and middle panel: Power spectra for OAM bases with TC ranges [-12, 12] (step size 1) and [-24, 24] (step size 2), showing maximum crosstalk values of -11.42 dB and -13.38 dB, respectively. Right panel: Statistic comparison of transmission performance (average power spectrum and crosstalk) for both bases. **b**, Encoding scheme: 24-bit RGB data modulated onto 24 OAM channels grouped by TC ranges [-24, -10], [-8, 8], and [10, 24]. Right panel shows an example multiplexed input field and the corresponding CPDP-SCH reconstruction. **c**, Decomposition power spectrum of the recovered field in b, confirming the encoded OAM components. **d**, Original 100× 100-pixel color test image and the reconstructed image with an error rate of 0.039%.

## 3.3 OAM reconstruction through dynamic scattering media

To evaluate the performance in dynamic conditions, we employed a motorized rotating ground glass as a controlled dynamic scattering medium (Fig. 5a). Fig. 5b displays the reconstructed amplitude and phase for TC values of 1, 7, 14 and 20 at different time points under a rotation speed of 20 revolutions per minute (r/min), confirming high fidelity despite medium dynamics.

We then quantitatively assessed the impact of rotational speed on the recovery of a vortex beam with TC=10. Power spectra for step-size-2 OAM basis were computed from 100 consecutive speckle pattern captures at each speed (Fig. 5c; see Supplementary Note 5 for detailed analysis). Fig. 5d shows the average power spectrum across the step-size-2 OAM basis for 100 measurements under a rotation speed of 20 r/min. The average power spectrum value and average crosstalk for single vortex beams transmission were 0.9372 and -12.02 dB, respectively, indicating that the recovered vortex beams retain good orthogonality even under dynamic scattering.

Finally, we transmitted four 100×100-pixel color images (flower, architecture, mountain, and penguins) pixel by pixel using OAM-multiplexing encoding at a rotation speed of 20 r/min. The reconstructed images and their corresponding transmission accuracies are shown in Fig. 5e. An average accuracy of 98.97% was achieved, confirming the robustness of CPDP-SCH for high-capacity OAM-multiplexed transmission in dynamic scattering environments.

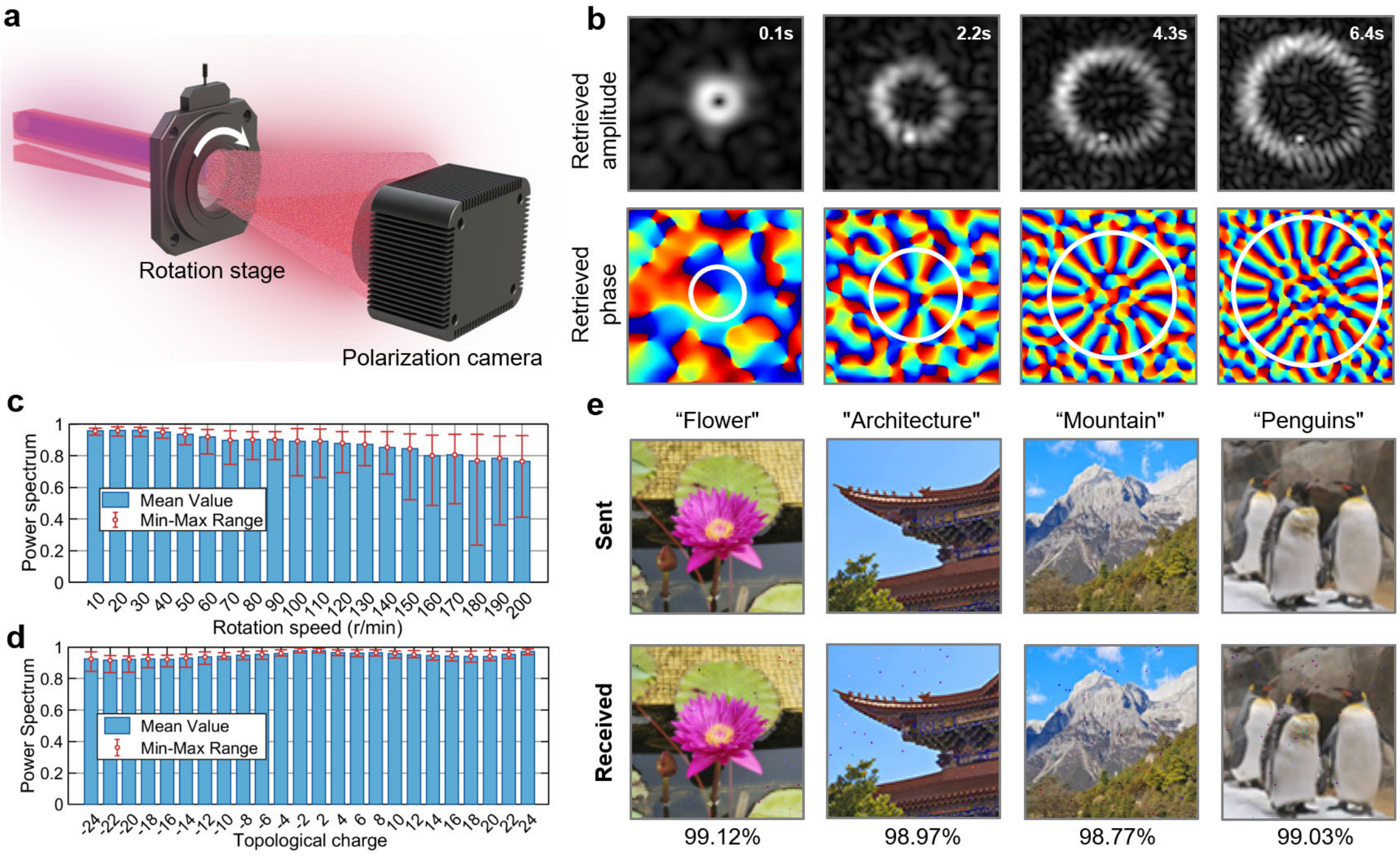

**Fig. 5**. **Dynamic scattering performance evaluation. a**, Experimental setup with a motorized rotation stage to generate controlled dynamic scattering via a rotating ground glass. **b**, Reconstructed amplitude (top row) and phase (bottom row) for vortex beams with TC = 1, 7, 14 and 20 at different time points under a rotation speed of 20 r/min. **c**, Power spectrum values of the reconstructed OAM mode (TC = 10) as a function of ground-glass rotation speed. **d**, Average power spectrum across the step-size-2 OAM basis (rotation speed of 20 r/min, 100 measurements). **e**, Top row: Four 100×100-pixel 24-bits color

images used as the transmission objects. Bottom row: Reconstructed images using OAM-multiplexing combined with CPDP-SCH under dynamic scattering. Numbers indicate the corresponding transmission accuracy.

## 4 Discussion and Conclusion

The experimental results demonstrate that the CPDP-SCH method achieves high-accuracy transmission of both single vortex beams and OAM-multiplexed fields through static and dynamic scattering media, enabled by the principle of correlation invariance. By leveraging this invariance together with polarization multiplexing, CPDP-SCH substantially enhances the information capacity of single-shot speckle measurements, allowing direct retrieval of the input field's complex amplitude without phase-shifting operations or iterative algorithms. This intrinsic feature of correlation invariance eliminates the need for pre-calibrating transmission matrices, collecting large training datasets, or performing repeated speckle calibrations for different vortex beams. Consequently, the method excels in dynamic scattering environments, where it effectively overcomes the severe accuracy degradation and decoding failures typically of conventional techniques. By harnessing correlation invariance, CPDP-SCH significantly broadens the scope of OAM-based optical communication and creates promising opportunities for high-capacity optical transmission and speckle-based encryption [46-49].

Beyond robustness to dynamic scattering, the CPDP-SCH implementation is remarkable compact. A single SLM combined with polarization optics achieves spatial-polarization multiplexing, while a polarization camera enables single-shot acquisition of dual-polarization off-axis speckle holograms. This eliminates the need for multi-detector setups or repeated acquisitions. Compared to traditional speckle correlation systems that require long-path off-axis reference beams, our method generates polarization-separated reference beams directly from distinct sub-regions of the SLM loaded with designed lens phase patterns, removing the need for external references. Critically, the object and reference beams for each polarization channel share identical transmission paths, enhancing system stability and perturbation resistance for reliable hologram generation and information recovery.

To broaden CPDP-SCH's applicability and further improve the dynamic-scattering transmission accuracy, future work will focus on two aspects. First, since the reference speckle field is currently generated using a lens-phase beam focused onto the scattering medium, optimal reconstruction typically requires the medium to be positioned near the focal plane. Replace this lens phase pattern with an axicon phase pattern [50-53] or a specially designed phase field [54-57] could generate Bessel beams with extended depth of focus, relaxing the axial-position constraints and enabling high-accuracy recovery during continuous axial displacement. Second, rapid rotation of the scattering medium slightly increases OAM-basis crosstalk relative to static conditions (Fig. 5; reducing OAM-multiplexing transmission accuracy). This stems from shorter integration times per speckle state at higher speeds, which limits the information captured per frame. Increasing laser power and/or using detectors with higher quantum efficiency could minimize this crosstalk. Simultaneously, replacing the liquid crystal-based SLM with a high-speed digital micromirror device (DMD; 23 kHz) combined with Lee-hologram complex modulation [58] would enable rapid generation of OAM-encoded information, facilitating faster, higher-capacity, and high-fidelity transmission through dynamic scattering media.

In conclusion, we have introduced the CPDP-SCH method, which exploits the concept of correlation invariance to achieve scattering-immune, high-fidelity OAM-encoded information transmission. This invariance, realized through cross-correlation of orthogonally polarized speckle holograms, cancels dynamic scattering phases while preserving deterministic object information. Implemented via a compact single-SLM design, the method achieves single-shot, high-accuracy OAM recovery without pre-calibration or training. Experimentally, CPDP-SCH achieved 99.61% accuracy for 24-bit OAM-multiplexed data transmission through static scattering media and maintained 98.97% accuracy under dynamic scattering conditions. This work overcomes a fundamental barrier for OAM-based systems and establishes a robust platform for high-capacity optical communications, optical encryption, and imaging in dynamic scattering environments.

## *Disclosures*

The authors declare no conflicts of interest.

## *Acknowledgments*

This work was supported by National Natural Science Foundation of China (82330061), Hong Kong Research Grant Council (15125724), Hong Kong Innovation and Technology Commission (MHP/206/24), Shenzhen Science and Technology Innovation Commission (JCYJ20220818100202005), Guangzhou Science and Technology Projects (2025B01J3013, 2025B03J0097), and Hong Kong Polytechnic University (P0045762, P0049101, P0048314, P0059222, P0054249).

## *Code and Data Availability*

The code and data underlying the results presented in this paper are not publicly available at this time but can be obtained from the authors upon reasonable request.

# Supplementary Material for: "Correlation invariance unlocks robust calibration-free OAM multiplexing transmission under dynamic scattering scenarios"

**Haoran Li,[a,b,†] Zhiyuan Wang,[a,b,†] Zhipeng Yu,[a,b,d,†] Xingpeng Du,[c] Tianting Zhong,[a,b] Jixiong Pu,[c] Ziyang Chen,[c] Vinu R V,[c,*] Xiangping Li,[d,*] and Puxiang Lai,[a,b,e,*]**

[a] Department of Biomedical Engineering, The Hong Kong Polytechnic University, Hong Kong, China

[b] Shenzhen Research Institute, The Hong Kong Polytechnic University, Shenzhen, China

[c] College of Information Science and Engineering, Huaqiao University, Xiamen, China

[d] Guangdong Provincial Key Laboratory of Optical Fiber Sensing and Communications, Institute of Photonics Technology, Jinan University, Guangzhou, China

[e] Photonics Research Institute, Hong Kong Polytechnic University, Hong Kong, China

[†] These authors contributed equally to the work.

[*] Correspondence to Vinu R V at vinurv@hqu.edu.cn, Xiangping Li at xiangpingli@jnu.edu.cn, and Puxiang Lai at puxiang.lai@polyu.edu.hk

### *Supplementary Note 1: Derivation of the Correlation Hologram*

In the implemented CPDP-SCH system, the composite beam generated by the SLM contains four fields: an OAM-encoded field, a plane wave, and two lens-phase fields, each assigned to a specific polarization. Horizontally polarized light carries the OAM-encoded field (single/multiplexed vortex beam) and one lens phase field; vertically polarized light carries a plane wave and the second lens-phase field. Each polarization component propagates independently through the scattering medium, and a polarization camera records the resultant speckle patterns. The optical propagation occurs in two stages: from the output modulation plane to the scattering medium plane, and from the scattering medium plane to the detection plane.

The free-space diffraction kernel between arbitrary planes is:

$$h_Z(x_2, x_1) \approx \frac{exp(ikZ)}{i\lambda Z} exp\left(ik\frac{|x_2 - x_1|^2}{2Z}\right), \tag{S1}$$

where $x_1$ and $x_2$ denote coordinates on the initial and final planes, respectively, $Z$

is the propagation distance, $\lambda$ is the wavelength of the incident light, and $k = \frac{2\pi}{\lambda}$ is the wavevector. Therefore, propagation to the scattering medium plane is described by:

$$\begin{cases} E_H(\rho_1) = \int h_f(\rho_1, \xi_1)\left(\sum_l E_{H,l}^{O}(\xi_1)\right)d\xi_1 = \sum_l \int h_f(\rho_1, \xi_1)E_{H,l}^{O}(\xi_1)d\xi_1 = \sum_l E_{H,l}^{O(f)}(\rho_1) \\ E_V(\rho_2) = \int h_f(\rho_2, \xi_2)E_0 d\xi_2 = E_0' \\ E_{RH}(\rho_1) = \int h_f(\rho_1, \xi_1)\, exp\left(-\frac{ik}{2f}|\xi_1|^2\right)d\xi_1 = circ\left(\frac{\rho_1 - \rho_s}{a}\right) \\ E_{RV}(\rho_2) = \int h_f(\rho_2, \xi_2) exp\left(-\frac{ik}{2f}|\xi_2|^2\right)d\xi_2 = circ\left(\frac{\rho_2 - \rho_s}{a}\right) \end{cases}, (S2)$$

where $\xi_1$ and $\xi_2$ denote the coordinates corresponding to the horizontal and vertical polarization directions, respectively, on the output modulation plane, while $\rho_1$ and $\rho_2$ are the corresponding coordinates on the scattering medium plane. Subscripts H and V indicate the horizontal and vertical polarization states of the respective field. The OAM-encoded beam at output modulation plane, formed by superposing vortex beams with different TCs, is expressed as $\sum_l E_{H,l}^{O}(\xi_1) = \sum_l \left(\frac{\sqrt{2}|\xi_1|}{\omega_0}\right)^{|l|} exp\left(-\frac{|\xi_1|^2}{{\omega_0}^2}\right) exp(il\varphi)$, where $l$ denotes the TC, $\omega_0$ is the beam waist, and $\varphi$ is the azimuthal angle. The corresponding field of the OAM-encoded beam propagating to the scattering medium plane is given by $E_H(\rho_1) = \sum_l E_{H,l}^{O(f)}(\rho_1)$, representing the superposition of individual vortex beams after propagating independently over distance $f$. Since vortex beams maintain their characteristic properties during free-space propagation, field $E_{H,l}^{O(f)}(\rho_1)$ at the scattering medium plane is also a vortex beam and can hence be expressed as $E_{H,l}^{O(f)}(\rho) = \int h_f(\rho_1, \xi_1)E_{H,l}^{O}(\xi_1)d\xi_1$. The plane wave at the output modulation plane is denoted by $E_0$, and its corresponding field at the scattering medium plane is also a plane wave, expressed as $E_V(\rho_2) = E_0'$. The two lens-phase fields term $exp\left(-\frac{ik}{2f}|\xi|^2\right)$ focus to identical positions $\rho_s$ on the scattering medium plane, forming circular apertures of diameter $a$.

After passing through the scattering medium and propagating a distance of $z_0$, the fields produce polarization-separate off-axis speckle holograms on the detection plane. The corresponding propagation process can be expressed as:

$$\begin{cases} E_H(r_1) = \int h_{z_0}(r_1,\rho_1)\left[E_H(\rho_1)e^{i\varphi(\rho_1)}\right] d\rho_1 \\ E_V(r_2) = \int h_{z_0}(r_2,\rho_2)\left[E_V(\rho_2)e^{i\varphi(\rho_2)}\right] d\rho_2 \\ E_{RH}(r_1) = \int h_{z_0}(r_1,\rho_1)\left[circ\left(\frac{\rho_1-\rho_s}{a}\right)e^{i\varphi(\rho_1)}\right] d\rho_1 \\ E_{RV}(r_2) = \int h_{z_0}(r_2,\rho_2)\left[circ\left(\frac{\rho_2-\rho_s}{a}\right)e^{i\varphi(\rho_2)}\right] d\rho_2 \end{cases}, \quad (S3)$$

where $r_1$ and $r_2$ denote the coordinates on the detection plane corresponding to the horizontal and vertical polarization directions, respectively; $E_H(r_1)$ and $E_V(r_2)$ represent the polarization-specific speckle patterns generated by the OAM-encoded beam and plane wave, respectively; $E_{RH}(r_1)$ and $E_{RV}(r_2)$ denote the speckle patterns formed by the lens-modulated fields in their respective polarization channels, serving as off-axis reference speckles for forming the corresponding off-axis speckle holograms. $\varphi(\rho)$ denotes the random phase modulation imposed by the scattering medium. Furthermore, the holograms $U_H$ and $U_V$, can be expressed as:

$$\begin{cases} U_H(r_1) = E_H(r_1) + E_{RH}(r_1) \\ U_V(r_2) = E_V(r_2) + E_{RV}(r_2) \end{cases}. \quad (S4)$$

The polarization camera captures the intensities of these two fields: $I_H(r_1) = |U_H(r_1)|^2$ and $I_V(r_2) = |U_V(r_2)|^2$, which is expressed in Fig.1b are the speckle holograms with horizontal and vertical polarization directions, respectively. The subregions show the off-axis fringe generated in the corresponding holograms.

For a random process obeying Gaussian statistics, the cross-correlation of the intensity fluctuations, $C(r_1,r_2)$, is given by:

$$C(r_1,r_2) = \langle \Delta I_H(r_1)\Delta I_V(r_2)\rangle = \left|c_{H,V}(r_1,r_2)\right|^2, \quad (S5)$$

where $< >$ denotes the ensemble average and $\Delta$ represents the intensity fluctuation relative to the mean intensity. $c_{H,V}(r_1,r_2)$ is the field cross-correlation function expressed by:

$$
\begin{aligned}
\left|c_{H,V}(r_1,r_2)\right|^2 &= \left|\left\langle \left(E_H(r_1)+E_{RH}(r_1)\right)\left(E_V(r_2)+E_{RV}(r_2)\right)^* \right\rangle\right|^2\\
&= |\langle E_H(r_1)E_V^*(r_2) + E_{RH}(r_1)E_V^*(r_2) + E_H(r_1)E_{RV}^*(r_2) + E_{RH}(r_1)E_{RV}^*(r_2)\rangle|^2\\
&= |\langle E_H(r_1)E_V^*(r_2) + E_{RH}(r_1)E_{RV}^*(r_2)\rangle|^2\\
&= |\langle E_H(r_1)E_V^*(r_2)\rangle + \langle E_{RH}(r_1)E_{RV}^*(r_2)\rangle|^2 \qquad .(S6)
\end{aligned}
$$

In Equation (6), the cross-terms $\langle E_{RH}(r_1)E_V^*(r_2)\rangle = 0$ and $\langle E_H(r_1)E_{RV}^*(r_2)\rangle = 0$ because the component in the pairs $E_{RH}(r_1), E_V^*(r_2)$ and $E_H(r_1), E_{RV}^*(r_2)$ originate from independent random modulations occurring in distinct regions of the scattering medium. Notably, the speckle field $E_H(r_1)$, generated by the input vortex beam or OAM-multiplexed beam after transmission through the scattering medium, is encoded within the final expression of $c_{H,V}(r_1,r_2)$.

Substituting $r+\Delta r$ for $r_1$ and $r$ for $r_2$, $C(r_1,r_2)$ transforms to:

$$
C(r_1,r_2) = |\langle E_H(r+\Delta r)E_V^*(r)\rangle + \langle E_{RH}(r+\Delta r)E_{RV}^*(r)\rangle|^2. \qquad (S7)
$$

Substituting Eq. (S1) and Eq. (S2) from the manuscript into Eq. (S7), the cross term within $C(r_1,r_2)$ containing the OAM-encoded information is:

$$
\begin{aligned}
\langle E_H(r+\Delta r)E_V^*(r)\rangle &= \iint h_{z_0}(r+\Delta r,\rho_1)h_{z_0}^*(r,\rho_2)\left[E_H(\rho_1)e^{i\varphi(\rho_1)}\right]\left[E_V^*(\rho_2)e^{-i\varphi(\rho_2)}\right]d\rho_1 d\rho_2\\
&= \iint h_{z_0}(r+\Delta r,\rho_1)h_{z_0}^*(r,\rho_2)\,e^{i(\varphi(\rho_1)-\varphi(\rho_2))}E_H(\rho_1)E_V^*(\rho_2)d\rho_1 d\rho_2 \qquad (S8)
\end{aligned}
$$

Incorporating the free-space propagation kernel (Eq. (1) in the manuscript) yields:

$$
\begin{aligned}
\langle E_H(r+\Delta r)E_V^*(r)\rangle &= \iint \frac{1}{\lambda^2 {Z_0}^2} exp\left(ik\frac{|r+\Delta r-\rho_1|^2}{2Z_0}\right) exp\left(-ik\frac{|r-\rho_2|^2}{2Z_0}\right) e^{i(\varphi(\rho_1)-\varphi(\rho_2))}E_H(\rho_1)E_V^*(\rho_2)d\rho_1 d\rho_2\\
&= \iint \frac{1}{\lambda^2 {Z_0}^2} exp\left(ik\frac{|r+\Delta r-\rho_1|^2-|r-\rho_2|^2}{2Z_0}\right) e^{i(\varphi(\rho_1)-\varphi(\rho_2))}E_H(\rho_1)E_V^*(\rho_2)d\rho_1 d\rho_2\\
&= exp\left(ik\frac{|r+\Delta r|^2-|r|^2}{2Z_0}\right)\iint \frac{1}{\lambda^2 {Z_0}^2} exp\left(ik\frac{\left(|\rho_1|^2-|\rho_2|^2\right)}{2Z_0}\right) exp\left(\frac{-ik\left((r+\Delta r)\rho_1 - r\rho_2\right)}{Z_0}\right)\\
&\quad * e^{i(\varphi(\rho_1)-\varphi(\rho_2))}E_H(\rho_1)E_V^*(\rho_2)d\rho_1 d\rho_2 \qquad (S9)
\end{aligned}
$$

On the one hand, the $r$-dependent phase factor outside the propagation integral, with $\Delta r$ has a very small value, rendering this phase factor canceled out during the estimation of the fourth-order correlation. On the other hand, since the coordinates $\rho_1$ and $\rho_2$ are on the same detection plane, it has $\rho_1 = \rho_2 = \rho$. Under this condition, Eq. (S2) can be simplified to:

$$\langle E_H(r+\Delta r)E_V^*(r)\rangle = \frac{1}{\lambda^2 {Z_0}^2}\int E_H(\rho)E_V^*(\rho)exp\left(\frac{-ik(\Delta r\rho)}{Z_0}\right)d\rho\,. \quad (S10)$$

Substituting Eq. (2) from the manuscript yields the final expression for the cross term containing the OAM-encoded information:

$$c_{H,V}^{(O)}(\Delta r) = \langle E_H(r+\Delta r)E_V^*(r)\rangle = \frac{1}{\lambda^2 {Z_0}^2}\int \left[\sum_l E_{H,l}^{O(f)}(\rho)\right] exp\left(\frac{-ik(\Delta r\rho)}{Z_0}\right)d\rho\,. \quad (S11)$$

Similarly, the final expression for the cross term containing the off-axis reference beams in both polarization directions is:

$$c_{H,V}^{(R)}(\Delta r) = \langle E_{RH}(r+\Delta r)E_{RV}^*(r)\rangle = \frac{1}{\lambda^2 {Z_0}^2}\int circ\left(\frac{\rho-\rho_s}{a}\right) exp\left(\frac{-ik(\Delta r\rho)}{Z_0}\right)d\rho. \quad (S12)$$

The correlation hologram is thus:

$$C(r_1, r_2) = \left|c_{H,V}^{(O)}(\Delta r) + c_{H,V}^{(R)}(\Delta r)\right|^2. \quad (S13)$$

This equation shows that the cross-correlation of the two polarization-channel holograms yields a holographic interference pattern that encodes the complete complex field of the input OAM beam, free from the random scattering phase.

### *Supplementary Note 2: Complex fields generation and beam alignment*

#### 2.1 Complex field modulation via SLM

In this work, the modulation method described in Ref. [45] was employed to achieve complex field modulation using an SLM. The principle underlying this approach is that any complex amplitude field can be computationally encoded into a corresponding phase pattern. By performing a Fourier series expansion of the pattern-modulated field, the coefficient of the first-order term is found to be proportional to the amplitude of the original complex field. Consequently, isolating and extracting this first-order term enables accurate recovery of the corresponding complex amplitude field.

For a complex field represented as $E = Aexp(i\phi)$, where $\phi$ and $A$ represent the phase and normalized amplitude of the field, respectively, the encoded phase pattern can be expressed as:

$$\psi(\phi, A) = f(A)sin(\phi), \qquad (S14)$$

where the factor $f(A)$ within this expression will be determined in the subsequent derivation. The q-th order coefficient $c_q^A$ of the Fourier series expansion of $\psi(\phi, A)$ is given by:

$$c_q^A = J_q[f(A)], \qquad (S15)$$

where $J_q$ denotes the integer-order Bessel function. Since the first-order coefficient $c_1^A$ is proportional to the normalized amplitude A of the complex field, it has the relationship:

$$J_1[f(A)] = CA. \qquad (S16)$$

Here, the constant $C \cong 0.581$ corresponds to the maximum value of the first-order Bessel function $J_1$, which occurs when $f(A) \cong 1.84$. Therefore, the function f(A) can be determined through numerical inversion of equation (S8). The function $f(A)$ takes values within the interval $[0,1.84]$, and consequently, the phase pattern $\psi(\phi, A)$ ranges approximately within $[-1.84,1.84] \cong [-1.17\pi, 1.17\pi]$.

Therefore, the final encoded phase pattern is expressed by:

$$\psi(\phi, A) = J_1^{-1}[CA]sin\big(\phi + 2\pi(u_0x + v_0y)\big), \qquad (S17)$$

where $J_1^{-1}$ represents the numerical inversion of $J_1$ function. The parameters $u_0$ and $v_0$ denote the grating frequencies along the spatial coordinates $x$ and $y$, respectively. Their purpose is to enable the separation of different diffraction orders in the Fourier plane. For this work, values $u_0 = 0.25$ and $v_0 = 0$ were used.

## 2.2 Polarization alignment procedure

Alignment of the two orthogonally polarized beams involves three stages:

1) **Coarse alignment**: Adjust the reflection angles of mirrors M2 and M3 to achieve overlap between the horizontally polarized OAM field and the vertically polarized plane wave within the plane behind lens L2. Subsequently, modify the phase pattern loaded onto the SLM, replacing the original OAM-generating encoding phase with the plane-wave-generating encoding phase used on the right side. This modification causes both left and right sections of the SLM to generate identical modulated optical fields. Readjust mirrors M2 and M3 to achieve rough alignment of the two square-shaped plane wave fields.
2) **Fine spatial overlap**: Precisely tune the reflection angle of M2 and the tilt angle of the PBS to ensure exact spatial overlap of the two square-shaped plane wave fields on any plane within the region between lens L2 and the ground glass (GG).
3) **Polarization overlap**: Reload the original phase pattern onto the SLM. Observe the resulting field at the GG plane using a standard camera. Perform fine adjustments to the tilt angle of the PBS to bring the focal points generated by the lens phase in both polarization directions into coincidence. At this stage, the horizontally polarized OAM field should be fully contained within the vertically polarized plane wave, as demonstrated by the results captured with a standard camera shown in Fig. S1.

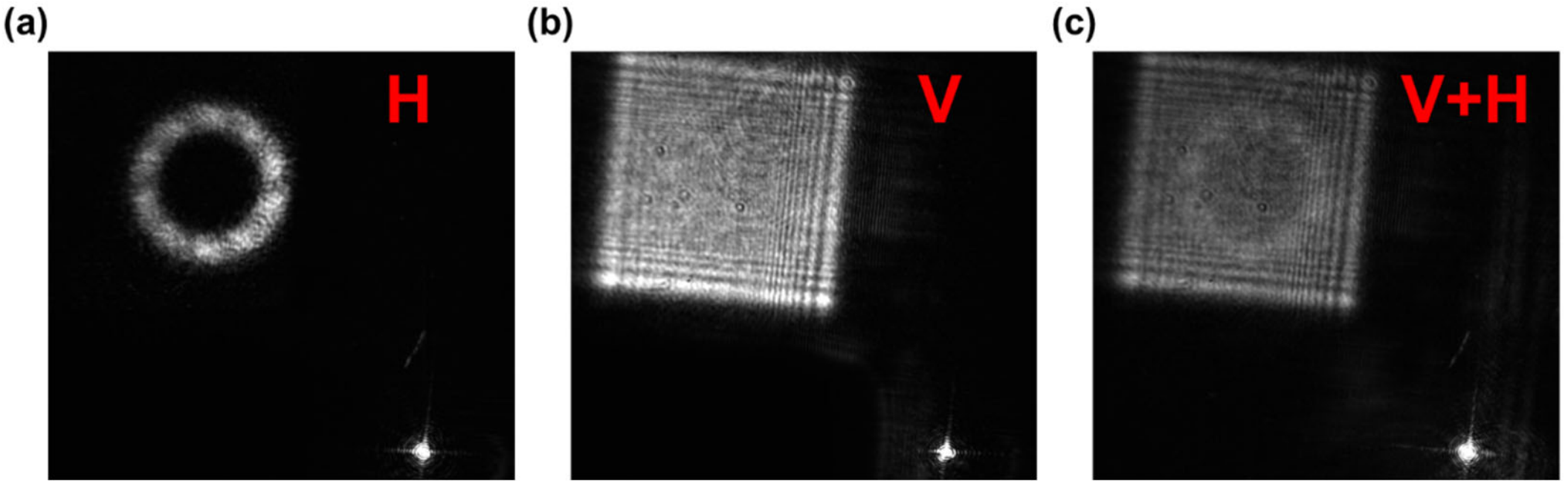

**Fig. S1**. **Intensity distribution of different polarization states on the ground glass (GG) plane after alignment.** The horizontally polarized OAM field (left) is fully overlapped by the vertically polarized plane wave (right), confirming precise spatial and polarization overlap for common-path transmission.

### *Supplementary Note 3: Digital reconstruction from the correlation hologram*

The CPDP-SCH method provides two orthogonally polarized off-axis speckle holograms. Their cross-correlation yields a correlation hologram that encapsulates the full complex information of the input OAM-encoded field. Reconstruction follows the standard off-axis holography workflow [42, 43] outlined in Fig. S2: First, compute the 2D Fourier transform of the correlation hologram; next, filter the +1 order spectrum (signal term) in the Fourier domain; then, center the filtered spectrum and apply an inverse Fourier transform; last, the result is the reconstructed complex field, from which amplitude and phase are extracted.

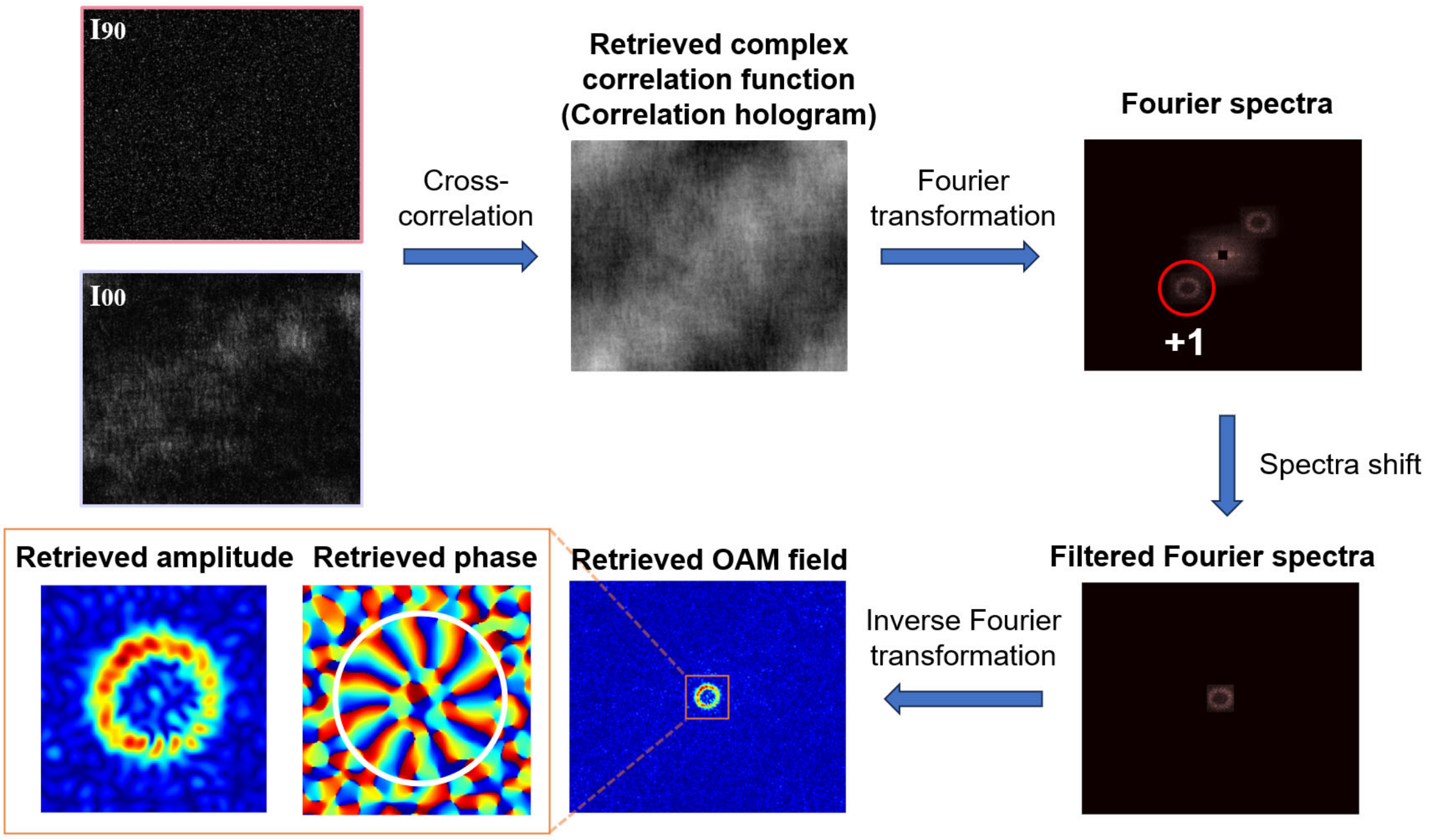

**Fig. S2**. **Flowchart for reconstructing the OAM-encoded field from the correlation hologram.** Following off-axis holography principles, Fourier transformation of the correlation hologram allows spectral filtering of the +1 order, which is then inverse-transformed to recover the complex field.

### *Supplementary Note 4: Off-axis parameter control*

In off-axis digital holography, the close proximity between the signal spectra and the central DC term in the Fourier plane results in information overlap, causing the filtered spectra to retain DC components that degrade reconstruction quality. CPDP-SCH allows flexible control of the off-axis separation using varying phase modulation patterns. In this study, the off-axis separation is determined by the relative distance on the GG plane between the focal points generated by the lens phase field and the OAM-encoded field. By switching phase patterns (Figs. S3a–c), the distance between the OAM-encoded field and the focal point on the GG plane can be increased, which correspondingly enhances the separation between the signal spectra and the DC term in the Fourier plane (Figs. S3e–g). Alternatively, adjusting the focal position relative to a fixed OAM-encoded field region on the GG plane using the corresponding phase modulation patterns (Figs. S3b, d) also increases the signal spectra–DC distance (Figs. S3f, h). Therefore, CPDP-SCH provides precise control over off-axis parameters and ensures high-quality reconstruction of the OAM-encoded field.

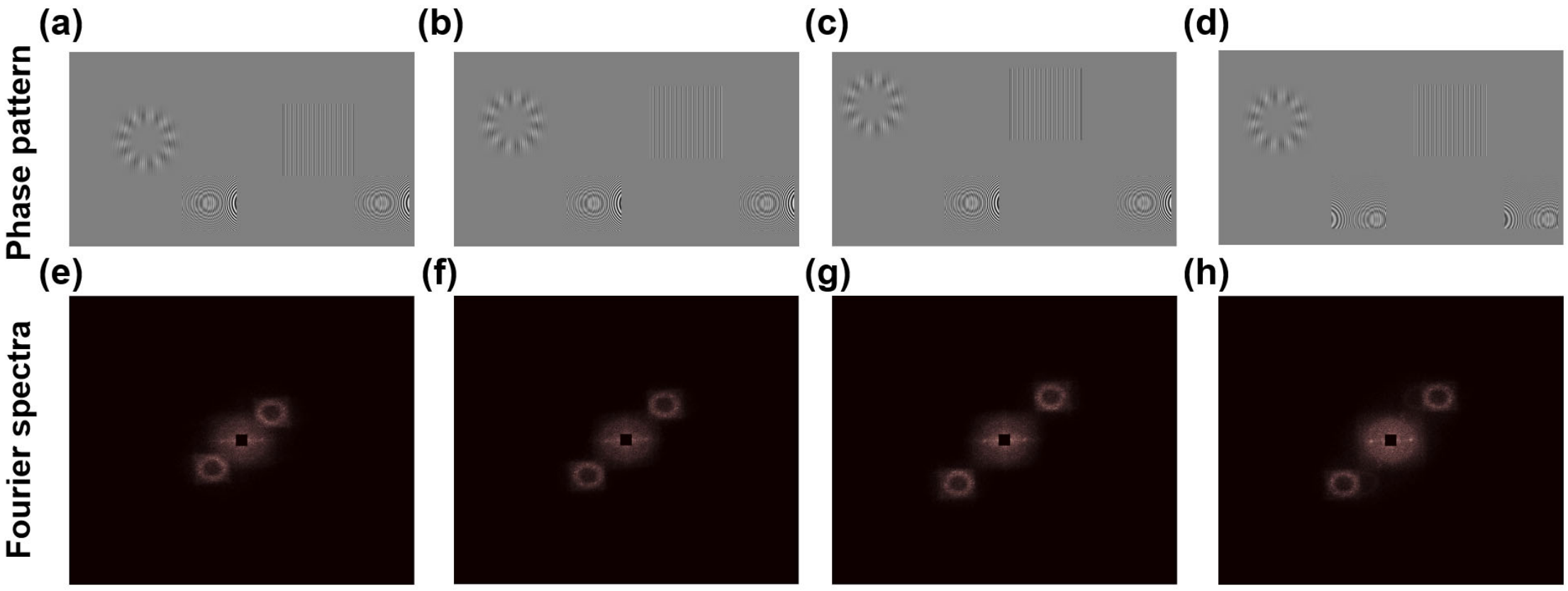

**Fig. S3**. **Off-axis parameter control via phase-pattern design.** (a–d) Phase patterns loaded on the SLM to vary the lateral separation between the OAM field and the focal spot on the GG plane. (e–h) Corresponding Fourier spectra of the correlation hologram; increased spatial separation shifts the signal spectrum farther from the DC term, avoiding overlap and improving reconstruction quality.

## *Supplementary Note 5: Scattering dynamics analysis*

To quantify the effect of scattering dynamics, we measured the recovery of a vortex beam (TC = 10) while rotating the ground glass at speeds from 10 to 200 r/min. For each speed, 100 consecutive speckle patterns were captured, and the power spectrum on the OAM basis was computed.

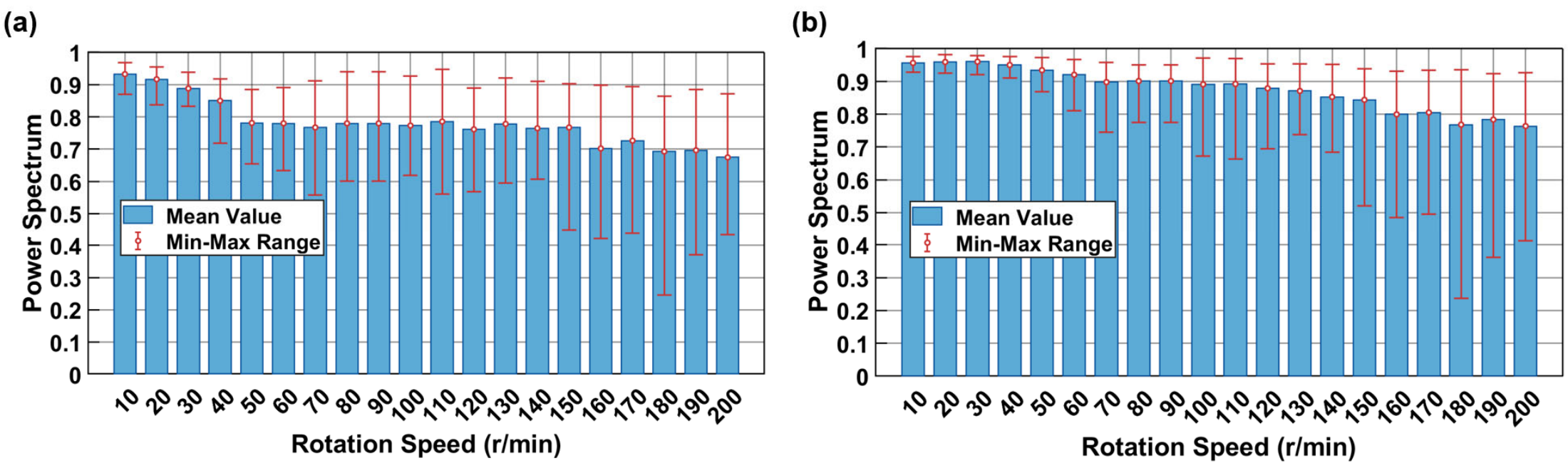

**Fig. S4**. **Rotation-speed dependence of OAM recovery accuracy.** Power-spectrum values of a recovered vortex beam (TC = 10) across rotation speeds (10–200 r/min) for OAM bases with (a) step size 1 and (b) step size 2. Each point represents the mean of 100 measurements. The step-size-2 basis exhibits slower decay, indicating greater robustness to dynamic scattering.

Figs. S4a-b show the power spectrum values across different rotation speeds within OAM bases with step sizes of 1 and 2, respectively. Due to the dynamic nature of the scattering medium during measurement and reconstruction, these values exhibit fluctuations at each speed. Notably, the power spectrum values decrease as the rotation speed increases for both step-size OAM bases. We further explore the underlying mechanisms responsible for this phenomenon in the following content. Furthermore, the power spectrum values in the step-size-2 OAM basis exhibit a lower decay rate and maintain a relatively high level as the rotation speed increases, indicating that the restored vortex beams in the step-size-1 OAM basis are more susceptible to crosstalk from adjacent OAM modes under significant dynamic changes. This crosstalk effect is comparatively reduced in the step-size-2 OAM basis, allowing it to preserve better transmission quality.

To understand this behavior, we analyzed the speckle patterns themselves. Figs. S5a-b present the Pearson correlation coefficient (PCC) between consecutively acquired speckle patterns $I_{90}$ at varying rotation speeds, and the speckle contrast of $I_{90}$ across these speeds, respectively. As shown in Fig. S5a, the PCC between two consecutively acquired speckle patterns remains below 0.025 at all rotation speeds, indicating that each speckle pattern essentially arises from an independent scattering event. Fig. S5b shows that speckle contrast decreases progressively with increasing rotation speed, attributable to the integration of a larger number of distinct speckle configurations over the camera exposure time at higher speeds. Consequently, the incoherent superposition of speckle patterns from multiple scattering states results in reduced contrast in the acquired speckle pattern.

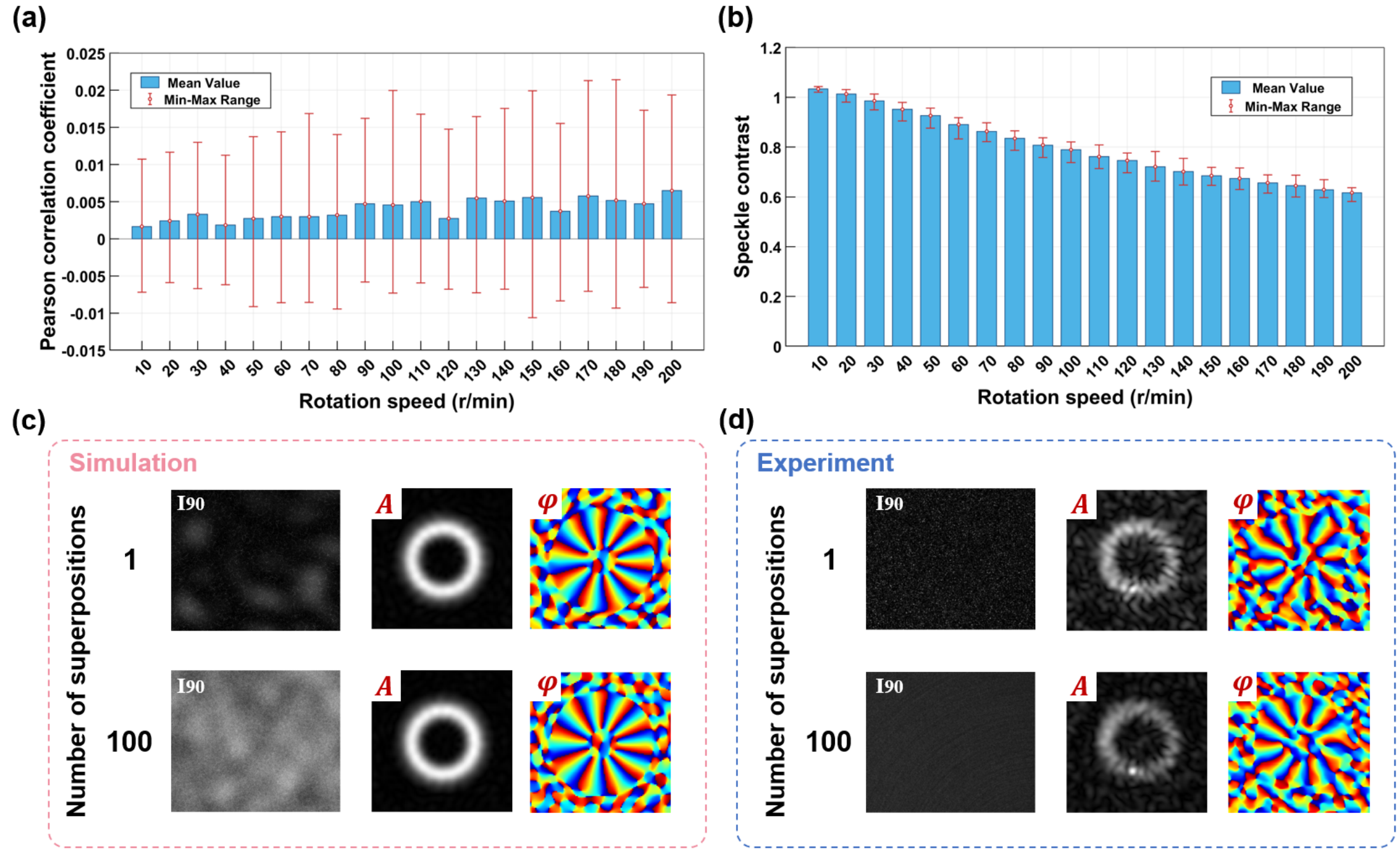

**Fig. S5. Speckle statistics and superposition invariance under dynamic scattering.** (a) Pearson correlation coefficient between consecutively captured speckle patterns ($I_{90}$) remains near zero, confirming independent scattering states. (b) Speckle contrast decreases with rotation speed due to integration of multiple states within the camera exposure. (c, d) Simulated and experimental reconstruction results from a single speckle and from the superposition of 100 speckles, demonstrating that reconstruction quality is preserved despite reduced contrast.

However, this incoherent superposition does not impair reconstruction quality. Fig. S5c and d compare the reconstruction quality of the OAM field obtained from a single speckle pattern and from the superposition of 100 distinct speckle patterns under simulated and experimental conditions, respectively. Although the speckle contrast after superposition is markedly lower than that of a single speckle pattern, both simulation and experimental results demonstrate that the final reconstruction quality remains intact and accurately reflects the input OAM characteristics.

This phenomenon can be explained as follows. The cross-correlation result $C(r_1, r_2)$ for the superposed speckle patterns can be expressed as:

$$C(r_1,r_2) = \langle \Delta\left(\sum_{i=1}^{n} I_{Hi}(r_1)\right)\Delta\left(\sum_{j=1}^{n} I_{Vj}(r_2)\right)\rangle = \sum_{i=1}^{n}\sum_{j=1}^{n}\langle \Delta I_{Hi}(r_1)\Delta I_{Vj}(r_2)\rangle, \quad (S18)$$

where $n$ denotes the number of superposed speckle patterns. Since each superposed speckle pattern originates from an independent scattering process, the random phases introduced by scattering are also independent. Consequently, the cross-correlation between $I_{Hi}(r_1)$ and $I_{Vj}(r_2)$ from different states is approximately zero:

$$\sum_{i\neq j}^{n}\langle \Delta I_{Hi}(r_1)\Delta I_{Vj}(r_2)\rangle = 0. \quad (S19)$$

Thus, the final result simplifies to the summation of the cross-correlations from individual speckle states:

$$C(r_1,r_2) = \sum_{i=j}^{n}\langle \Delta I_{Hi}(r_1)\Delta I_{Vj}(r_2)\rangle. \quad (S20)$$

Since the input field remains unchanged, the cross-correlation result for each individual speckle pattern is approximately identical. Therefore, the final reconstruction result obtained from the superposed speckle patterns is equivalent to that obtained from a single speckle pattern.

We subsequently analyzed the information content of speckle patterns acquired at varying rotation speeds. Each acquisition captures two off-axis speckle holograms corresponding to orthogonally polarized components, allowing examination of their spectral information in the Fourier plane. Figure S6 presents the results across different rotation speeds. As the rotation speed increases, the intensities of the plane wave and OAM field components associated with the vertically and horizontally polarized off-axis speckle holograms in the Fourier plane progressively diminish, eventually becoming nearly negligible. Consequently, the signal spectra of the calculated correlation hologram in the Fourier plane also decreases, leading to degraded amplitude and phase reconstruction at higher rotation speeds.

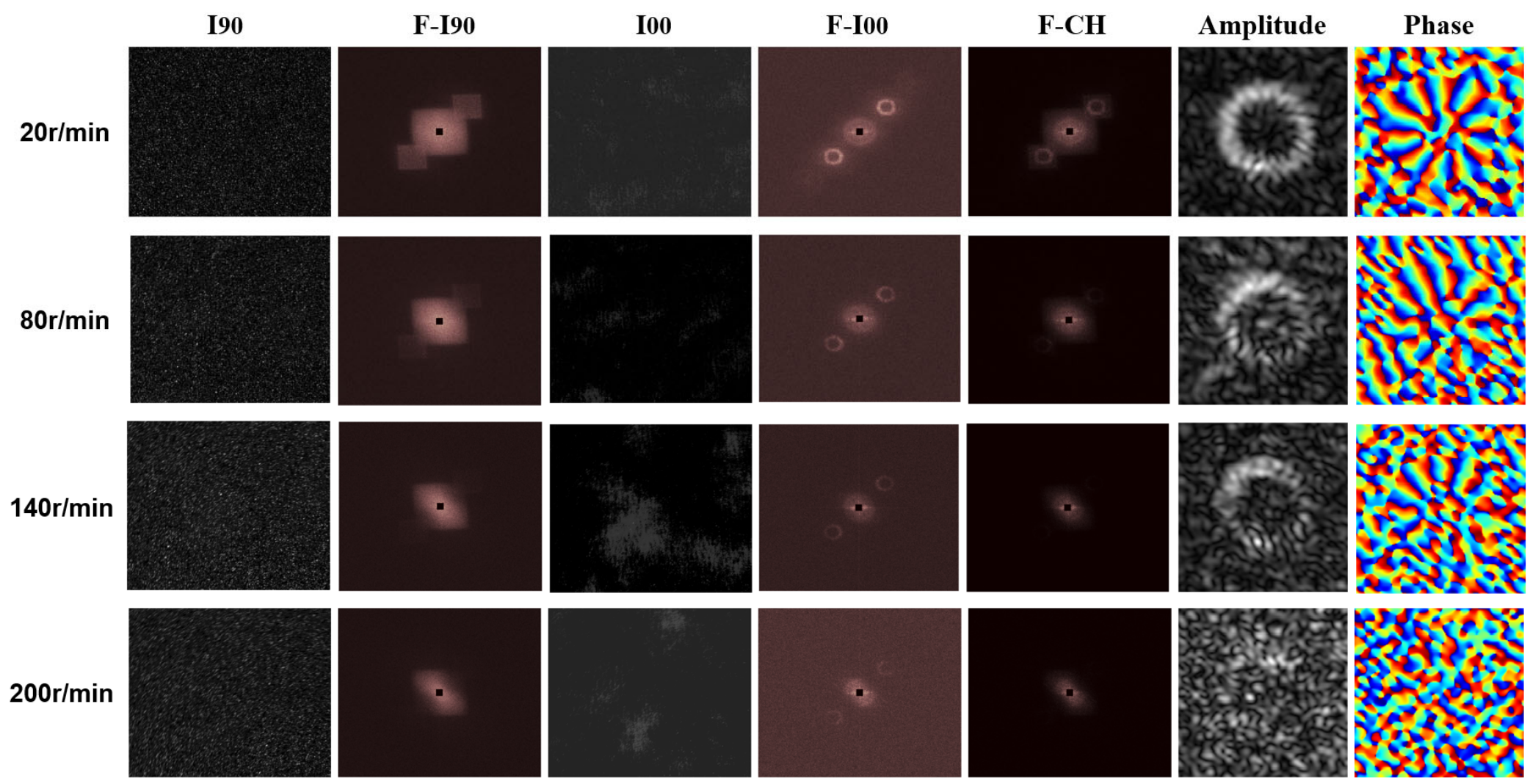

**Fig. S6**. **Signal degradation at high rotation speeds.** Columns show (1,3) raw speckle patterns ($I_V$, $I_H$), (2,4) their Fourier spectra, (5) Fourier spectrum of the correlation hologram, and (6,7) recovered amplitude and phase of the OAM field, for increasing rotation speeds (top to bottom). At higher speeds, the signal strength in the Fourier domain diminishes because the camera captures only a fraction of the incident field per exposure, yet the OAM information remains recoverable.

This phenomenon can be explained as follows: For a fixed camera exposure time, higher rotation speeds result in extremely short integration times for each distinct speckle state during rotation, causing the camera to capture only a limited portion of the information from the incident plane wave and the OAM-encoded field. This insufficient recording impairs accurate reconstruction of the correlation hologram, thereby compromising the fidelity of the recovered results. Therefore, in practical applications of CPDP-SCH, it is crucial to maintain reasonable laser output power and/or utilize detectors with higher quantum efficiency. These measures ensure that, even under rapid scatterer variation, a substantial portion of the incident field information is effectively recorded and preserved within the speckle patterns, ultimately enabling high-fidelity information reconstruction.